\documentclass[submitting]{nst}

\usepackage{subfigure,dcolumn}
\usepackage{epstopdf}
\usepackage{mhchem}
\usepackage{upgreek}
\usepackage{comment}
\usepackage{xspace}
\usepackage{placeins}
\usepackage{relsize}
\usepackage{multirow}

\usepackage{listings}
\newcommand{\rmnum}[1]{\romannumeral #1}
\newcommand{\pbpb}           {Pb--Pb\xspace}

\newcommand{\kT}           {\ensuremath{k_{\rm T}}\xspace}

\newcommand{\kstar}        {\ensuremath{k^*}\xspace}
\newcommand{\rstar}        {\ensuremath{r^*}\xspace}
\newcommand{\mt}           {\ensuremath{m_{\rm{T}}}\xspace}
\newcommand{\kt}           {\ensuremath{k_{\rm{T}}}\xspace}

\newcommand{\rcore}           {\ensuremath{R_{\rm{core}}}\xspace}

\newcommand{\seven}        {$\sqrt{s}~=~7$~Te\kern-.1emV\xspace}
\newcommand{\thirteen}        {$\sqrt{s}~=~13$~Te\kern-.1emV\xspace}
\newcommand{\tev}          {Te\kern-.1emV\xspace}

\newcommand{\Led}         {Lednick\'y--Lyuboshits\xspace}

\newcommand{\pP}           {\ensuremath{\mathrm{p}\mbox{--}\mathrm{p}}\xspace}

\newcommand{\pipi}          {\ensuremath{\pi\mbox{--}\pi}\xspace}
\newcommand{\pipiP}          {\ensuremath{\pi^{+}\mbox{--}\pi^{+}}\xspace}
\newcommand{\pipiN}          {\ensuremath{\pi^{-}\mbox{--}\pi^{-}}\xspace}
\newcommand{\Kp}           {\ensuremath{\mathrm{K}\mbox{--}\mathrm{p}}\xspace}

\newcommand{\pLpair} {\ensuremath{\mathrm{p}\mbox{-}\Lambda}\xspace}

\newcommand{\UGeVc}{GeV/$\it{c}$\xspace}

\begin{document}

\title{Investigating the pion emission source in pp collisions using the AMPT model with sub-nucleon structure}
\thanks{This work is supported by the National Natural Science Foundation of China (Nos. 12061141008, 12147101, and 12322508) and the Science and Technology Commission of Shanghai Municipality (23590780100). LZ acknowledges the support of the Fundamental Research Funds for the Central Universities, China University of Geosciences (Wuhan) with No. G1323523064.}

\author{Dong-Fang Wang}
\affiliation{Key Laboratory of Nuclear Physics and Ion-beam Application (MOE), Institute of Modern Physics, Fudan University, Shanghai 200433, China}
\affiliation{Shanghai Research Center for Theoretical Nuclear Physics, NSFC and Fudan University, Shanghai 200438, China}

\author{Mei-Yi Chen}
\affiliation{Key Laboratory of Nuclear Physics and Ion-beam Application (MOE), Institute of Modern Physics, Fudan University, Shanghai 200433, China}
\affiliation{Shanghai Research Center for Theoretical Nuclear Physics, NSFC and Fudan University, Shanghai 200438, China}

\author{Yu-Gang Ma}
\email[]{mayugang@fudan.edu.cn}
\affiliation{Key Laboratory of Nuclear Physics and Ion-beam Application (MOE), Institute of Modern Physics, Fudan University, Shanghai 200433, China}
\affiliation{Shanghai Research Center for Theoretical Nuclear Physics, NSFC and Fudan University, Shanghai 200438, China}

\author{Qi-Ye Shou}
\email[]{shouqiye@fudan.edu.cn}
\affiliation{Key Laboratory of Nuclear Physics and Ion-beam Application (MOE), Institute of Modern Physics, Fudan University, Shanghai 200433, China}
\affiliation{Shanghai Research Center for Theoretical Nuclear Physics, NSFC and Fudan University, Shanghai 200438, China}

\author{Song Zhang}
\affiliation{Key Laboratory of Nuclear Physics and Ion-beam Application (MOE), Institute of Modern Physics, Fudan University, Shanghai 200433, China}
\affiliation{Shanghai Research Center for Theoretical Nuclear Physics, NSFC and Fudan University, Shanghai 200438, China}

\author{Liang Zheng}
\email[]{zhengliang@cug.edu.cn}
\affiliation{School of Mathematics and Physics, China University of Geosciences (Wuhan), Wuhan 430074, China}
\affiliation{Shanghai Research Center for Theoretical Nuclear Physics, NSFC and Fudan University, Shanghai 200438, China}

\begin{abstract}

The measurement of momentum correlations of identical pions serves as a fundamental tool for probing the space-time properties of the particle emitting source created in high-energy collisions. Recent experimental results have shown that, in pp collisions, the size of the one-dimensional primordial source depends on the transverse mass (\mt) of hadron pairs, following a common scaling behavior, similar to that observed in Pb--Pb collisions. 
In this work, a systematic study of the \pipi source function and correlation function is performed using the multiphase transport model (AMPT) to understand the properties of the emitting source created in high multiplicity pp collisions at $\sqrt{s}=13$ TeV. 
The \mt scaling behavior and pion emission source radii measured by ALICE experiment can be well described the model with sub-nucleon structure. These studies shed new light on the understanding of the effective size of the \pipi emission source and on studying the intensity interferometry in small systems with a transport model.

\end{abstract}


\maketitle

\section{Introduction}\label{sec:intro}

The correlation of two particles at small relative momentum, known as femtoscopy, provides a unique method for directly probing the properties of particle emission and subsequent final-state interactions (FSI)~\cite{LisaReview}. 
To qualify the strength of the correlation, the two-particle correlation function $C(\kstar)$, is defined theoretically using the Koonin--Pratt equation~\cite{LisaReview,LauraFreview}
 \begin{equation}
    C(\kstar) = \int S(\mathbf{r^*}) |\Psi(\mathbf{r^*},\mathbf{k^*})| \text{d}^3 r^*,
     \label{Eq:Koonin-Pratt}
 \end{equation}
where $r^*= |\mathbf{r}^*_1-\mathbf{r}^*_2|$ and $\kstar = |\mathbf{p}^*_1-\mathbf{p}^*_2|/2$ represent the relative distance and momentum between the two particles in the pair rest frame (denoted by $*$), respectively. By definition, $C(\kstar)$ consists of two main components: the emission source function $S(\mathbf{r}^*)$, which describes the probability of producing two particles at a relative distance $\mathbf{r}^*$, and the wave function $\Psi(\mathbf{r^*},\mathbf{k^*})$, which is the asymptotic form of the combination of outgoing plane wave and scattered wave~\cite{QMgriffiths1984}. Typically, assuming the emission source is known (e.g., an isotropic Gaussian), one can extract the interaction between two particles of interest, often through the \Led parameterization~\cite{Lednicky:1981su} quantified by the scattering length and effective range~\cite{LednickFinalSize}, which has been successfully applied in various measurements in heavy-ion collisions (HICs)~\cite{ppFemtoSTAR,alice276,ks0kPbALICE,kpPbALICE}.

Alternatively, with the known interaction, the spatial extent and the duration of the emission source can be investigated through the interference of identical particles (e.g., pions)~\cite{Pioninterf1979,Pioninterf1991}. Such an intensity interferometry in HICs is commonly known as Hanbury Brown–Twiss analysis (HBT)~\cite{BEC2010_pp900,BEC2011_PbPb276}. 
The range of the strong interaction between two charged pions is expected to be around 0.2 fm~\cite{BEC1988}, and the scattering length $a_{0}^{I=2}$ is $-0.0444$ fm~\cite{pipiScatter,pipiEffectRange}, indicating that the effect of the strong interaction on the \pipi correlation function should be negligible. As a result, the \pipi interaction is primarily dominated by long-range Coulomb forces and the Bose–Einstein effect (quantum statistics). 

Femtoscopic studies show that, in both pp and \pbpb collisions, the source size decreases distinctly as a function of the pair's transverse mass \mt, defined as $\mt = \sqrt{\langle k_{\rm T} \rangle ^{2}+m^{2}}$, where $\kT=\left|\mathbf{p}_{\mathrm{T}, 1}+\mathbf{p}_{\mathrm{T}, 2}\right| / 2$ is the transverse momentum of the pair at rest frame and $m$ is the particle mass. This phenomenon, commonly referred to as \mt-scaling, has been observed for both identical mesons and baryons, such as pions, kaons, and protons~\cite{alice276,SourceMaxi}, and for non-identical particles, such as \pLpair~\cite{ppSource} and \Kp~\cite{SourceMaxi}. In \pbpb collisions, the scaling is typically attributed to the presence of the collective expansion of the system, i.e., radial flow~\cite{LisaReview}, which can be well described by the (3+1)D-dimensional hydrodynamic models~\cite{therm2014,therm2018,yu2019,Shapoval2014,THERMINATOR2}. In pp collisions, 
the range of the strong interaction ($\approx 1\mbox{--}2$ fm) is comparable to the source size ($1\mbox{--}3$ fm), 
and the hadronization is believed to occur on a similar timescale for all hadrons, which would lead to a corresponding \mt-scaling. On the other hand, unexpected strong collectivity has been experimentally observed in pp collisions in recent years~\cite{ppflow1,ppflow2,ppflow3,ppflow4}, and its origin remains not fully understood. Hence, the emission source and \mt-scaling in pp collisions have become even more intriguing and have gained considerable attention.

In addition to phenomenological models traditionally used in HICs for femtoscopic studies, such as EPOS~\cite{EPOSmodel}, UrQMD~\cite{UrQMD1,UrQMD2,li_effects_2022,li_transport_2022,fang_azimuthal-sensitive_2022,fang_simulation_2023}, HIJING~\cite{li_probing_2021}, CRAB~\cite{CRAB} and others~\cite{Li2002AMPTpiSource,Li20083DAMPTpiSource,piSourceNICA,piSourceEPOS,UrQMDpiSource,pipiUrQMDHIC}, CECA~\cite{CECA} offers a novel numerical approach to investigate the emission source. However, a comprehensive description that reasonably aligns collective flow with femtoscopy remains incomplete in pp collisions, though the former is recognized as a driving force behind the latter. It should be particularly noted that, the collective flow in pp collisions can be successfully reproduced by a multi-phase transport model (AMPT) implementing the sub-nucleon geometry, as demonstrated in recent works~\cite{ZL3qAMPT,Xinlipp13Cumulants,zheng_disentangling_2024}. This configuration, which incorporates the constituent quark assumption for protons, can generate a large initial spatial eccentricity, consequently leading to a significant long range azimuthal correlation in pp collisions. Therefore, it is crucial and nature to further explore if such a framework is also valid for revealing space-time characteristics. This work presents the first attempt to model the correlation function, the emission source, and \mt-scaling in high-multiplicity pp collisions at $\sqrt{s}=13$~TeV using the state-of-the-art AMPT.

This paper is organized as follows: In Sec.~\ref{sec:method}, a short introduction to the model and key parameters is provided. This is followed by an overview of the femtoscopic methodology which includes the source function and the framework used for providing accurate FSI of pion pairs. In Sec.~\ref{sec:result}, the impact of various physical factors, such as the parton re-scattering cross section $\sigma_{p}$, initial partonic distribution, short-lived resonances, and hadron re-scattering processes on the emission source is examined. Most importantly, the \mt-dependence of the \pipi source size is investigated. Finally, a brief summary is provided in Sec.~\ref{sec:summary}.

\section{MODEL AND METHODOLOGY}\label{sec:method}
\subsection{AMPT model}

The AMPT hybrid dynamic model~\cite{AMPTorigin,lin_further_2021}, which includes both partonic and hadronic scatterings, has been extensively used to study various key features of HICs, such as hadron production~\cite{dong_study_2024,SciChinaJinS}, collectivity~\cite{tang_investigating_2024,WangHai,wang_calculation_2024,ma_effects_2024} and phase transition~\cite{chen_transport_2024}. In recent years, the model has also been extended to small systems, i.e., pp and p--Pb collisions~\cite{Xinlipp13Cumulants}.
AMPT consists of four key components to simulate the collision process: the initial conditions, generated using the Heavy Ion Jet Interaction Generator (HIJING) model~\cite{HIJING-1,HIJING-2}; the partonic interactions, described by Zhang's Parton Cascade (ZPC) model~\cite{ZPCModel}; the hadronization process, which occurs through either Lund string fragmentation or a coalescence model; and the hadronic rescatterings, modeled by A Relativistic Transport (ART) model~\cite{ARTModel}. The model has two versions: (1) the string melting version, in which a partonic phase is generated from excited strings in the HIJING model, and a simple quark coalescence model combines partons into hadrons; and (2) the default version, which proceeds only through a pure hadron gas phase. 

This work is based on the AMPT with the string melting configuration incorporating sub-nucleon geometry when sampling the initial transverse positions of parton sources before converting them into constituent quarks (denoted by ``3 quarks"). 
This special tune, introduced in Ref.~\cite{ZL3qAMPT}, is able to successfully reproduce the spectra and elliptic flow of identified hadrons in pp collisions at TeV scales. Details regarding the initial partonic distribution will be described in Sec.~\ref{sec:result} A.
To illustrate the effects of the parton rescattering process, the value of $\sigma_{\text{p}}$ in ZPC is set to 1.5 and 10 mb, with 1.5 mb typically applied for larger systems~\cite{HeQuarkCoalAMPT}.

High-multiplicity events in AMPT are selected based on the number of charged particles in the pseudorapidity regions $-3.7<\eta<-1.7$ and $2.8<\eta<5.1$, corresponding to the acceptance of ALICE V0 detector. Additionally, an average multiplicity of approximately 30 charged particles is considered in the pseudorapidity interval $|\eta|<0.5$, following the event classification scheme used in ALICE pp collisions~\cite{pOmegaCFinpp}.
Using the particle selection criteria from this ALICE measurement~\cite{SourceMaxi}, charged pions are selected in the pseudorapidity range $|\eta|<0.8$ within the transverse momentum ($p_{\rm T}$) range of $\mbox{0.14--4.0}$~\UGeVc in this work.

\subsection{The correlation function and emission source}
The correlation function is introduced in Eq.~\eqref{Eq:Koonin-Pratt}. In this work, assuming the emission source is identical in all spatial directions, a single scalar \kstar is considered instead of the general three-dimensional $\mathbf{k*}$. The source function $S(r^*)$ describes the probability of producing two particles at a relative distance \rstar and is commonly modeled with a Gaussian profile
\begin{equation}
S(r^{*})=\frac{1}{\left(2 \pi R_{ab}^2\right)^{3 / 2}}  \exp{\left(-\frac{r^{* 2}}{2R_{ab}^2}\right)}.
\label{Eq:gauss_source}
\end{equation}
Here, $R_{ab}$ represents the general expression for the two-particle source radius of the $a\mbox{-}b$ pair. For identical particle pairs (\( a = b \)), this simplifies to \( R_{aa} = \sqrt{2} R_{a} \).
Note that \rstar denotes the relative distance between particles in specific pairs, while $R$ typically represents the variance of the distribution, reflecting the overall characteristics of the \rstar distribution contributed by many pairs.
A two-component source, consisting of a core from primary particles and a halo formed by resonance decays, has been used to describe Bose-Einstein correlations between identical pions in HICs~\cite{BEsourceHalo}. This paper follows the same nomenclature. In additional, this ``resonance halo", arising from short-lived, strongly decaying resonances ($\it{c}\tau\leq$ 5 fm), significantly increases the source size by introducing exponential tails to the source function of \pP and \pipi, as observed in recent pp collisions measurements~\cite{ppSource,SourceMaxi}. 
The $R_\text{core}$ manifest core Gaussion source radius and $R_\text{eff}$ represent effective Gaussion source radii which contains resonances effect. For previous measurements in \pbpb~\cite{alice276}, $R_\text{inv}$ might also be used to manifest source radius of identical pairs, called single-particle source radius which same as $R_{a}$. In this paper, the observed core source radius of \pipi is expressed as $R_\text{core}$, with $R_\text{core}=R_{\pi}=R_{\pi\pi}/\sqrt{2}$. 
Previous studies~\cite{CMS2020pipi13TeV, ATLAS2015pipi, CMS2018pipi} did not explicitly account for the effect of resonances, instead using a Cauchy/Exponential-type source parameterization~\cite{BEC_LevySource},
\begin{equation}
S\left(r^*\right)=\frac{1}{\pi^{2}} \frac{R_{\exp }}{(R_{\exp }^2+r^{* 2})^{2}},
\label{Eq:cauchy_source}
\end{equation}
where the Cauchy source size is denoted by $R_{\exp}$. In the absence of angular dependence, the probability of emitting two particles at given $r^{*}$ can be obtained by a simple integration over the solid angle: $S_{4\pi}(r^{*}) = 4\pi r^{*2} S(r^{*})$. In a different perspective, Hanbury-Brown--Twiss (HBT) interferometry measurements~\cite{pipiHBTinpp2011,pipiHBTinPbPb2011} indicate that the shape of the correlation function obtained in the Longitudinally Co-Moving System (LCMS) are different in the $R_\text{long}$, $R_\text{side}$, and $R_\text{out}$ directions. 
However, for simplicity, results in this work is based on isotropic Gaussion source shown in Eq~\ref{Eq:gauss_source}. Due to the fundamental assumptions in the Lednick\'{y} parameterization~\cite{LednickFinalSize}, the effective range expansion of scattering amplitude is not valid for small systems, particularly in pp collisions~\cite{Bhawanipd}.Therefore, the two-particle wave function is obtained using the ``Correlation Analysis Tool using the Schr\"{o}dinger Equation" (CATS) framework~\cite{Dimi_CATS}, which numerically solves the Schr\"{o}dinger equation for a configurable interaction potential.
In this work, the phase space of charged particle (positions and momenta) is provided by the AMPT model, while the CATS framework is used to accurately account for the FSI of the pairs to construct the \pipi correlation function.

\begin{figure}
    \centering
    \includegraphics[width=0.5\textwidth]{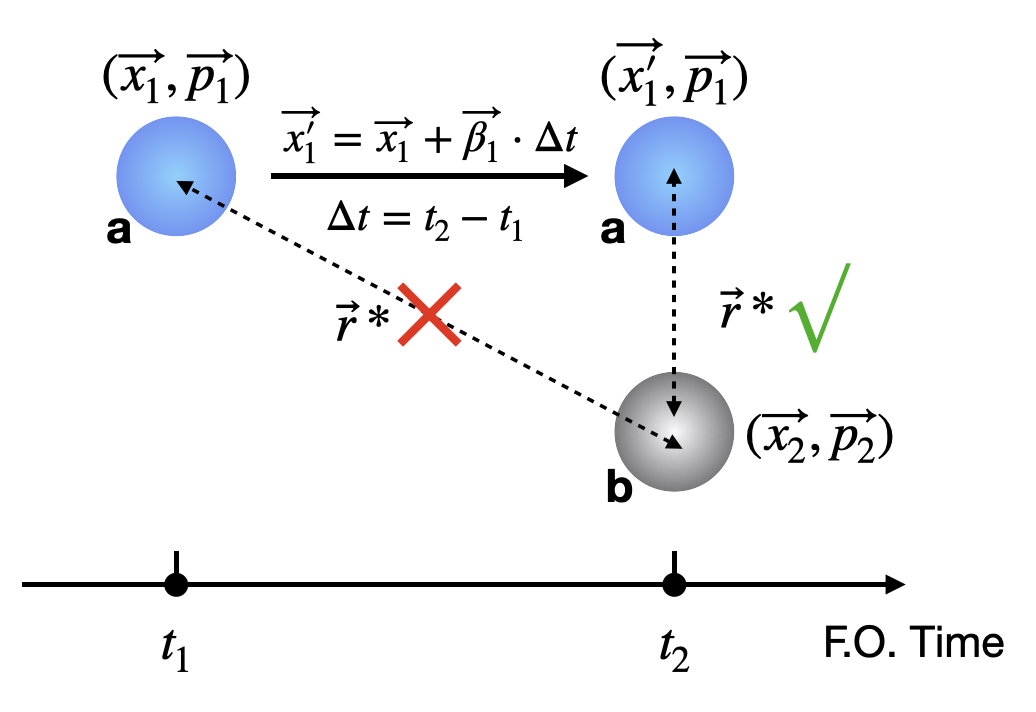}
    \caption{
    A sketch illustrates the modification of the coordinate $\vec{x_1}$ to $\vec{x'_1}$ for particle $a$ (blue disk), which is generated at time $t_1$, due to the different freeze-out time compared to particle $b$ (gray disk), generated at time $t_2$, in pairing the two particles based on the AMPT framework. The coordinate system is defined by the rest frame of the two particles and is consistent with Eq.~\eqref{Eq:Koonin-Pratt}, where $\vec{r^*}$ represents their relative distance (dash-dotted lines).}
    \label{fig:boostT}
\end{figure}

With a time step of 0.2 fm/$\it{c}$ in the AMPT computational framework, the particle generated earlier must propagate along its momentum for the time difference between the pair to satisfy the equal emission time condition, as illustrated in Fig.~\ref{fig:boostT}. Considering a pair of two particles, labeled as particle $a$ and $b$. The particle $a$, represented as a blue disk, is emitted at freeze-out (F.O.) time $t_{1}$ with position and momentum $(\vec{x_{1}}, \vec{p_{1}})$, earlier than the particle $b$, represented as a gray disk, which is emitted at time $t_{2}$ with $(\vec{x_{2}}, \vec{p_{2}})$. Under equal emission time condition, particle $a$ must propagate over a certain distance $\Delta \vec{x_{1}} = \vec{\beta_{1}}\cdot (t_{2}-t_{1})$ along the direction of its velocity $\vec{\beta_{1}}$. The distance of the pair $r^{*}$ is then calculated using $(\vec{x^{\prime}_{1}}, \vec{p_{1}})$ and $(\vec{x_{2}}, \vec{p_{2}})$, where $\vec{x^{\prime}_{1}}=\vec{x_{1}}+\Delta \vec{x_{1}}$.
In the model, no difference is expected between $\pi^{+}$ and $\pi^{-}$. Therefore, in the following text, the term \pipi refers to the combination of \pipiP and \pipiN pairs.

\section{results and discussion}\label{sec:result}
\subsection{Effect of initial partonic distribution}\label{sec:a}
\begin{figure}
    \centering
    \includegraphics[width=0.5\textwidth]{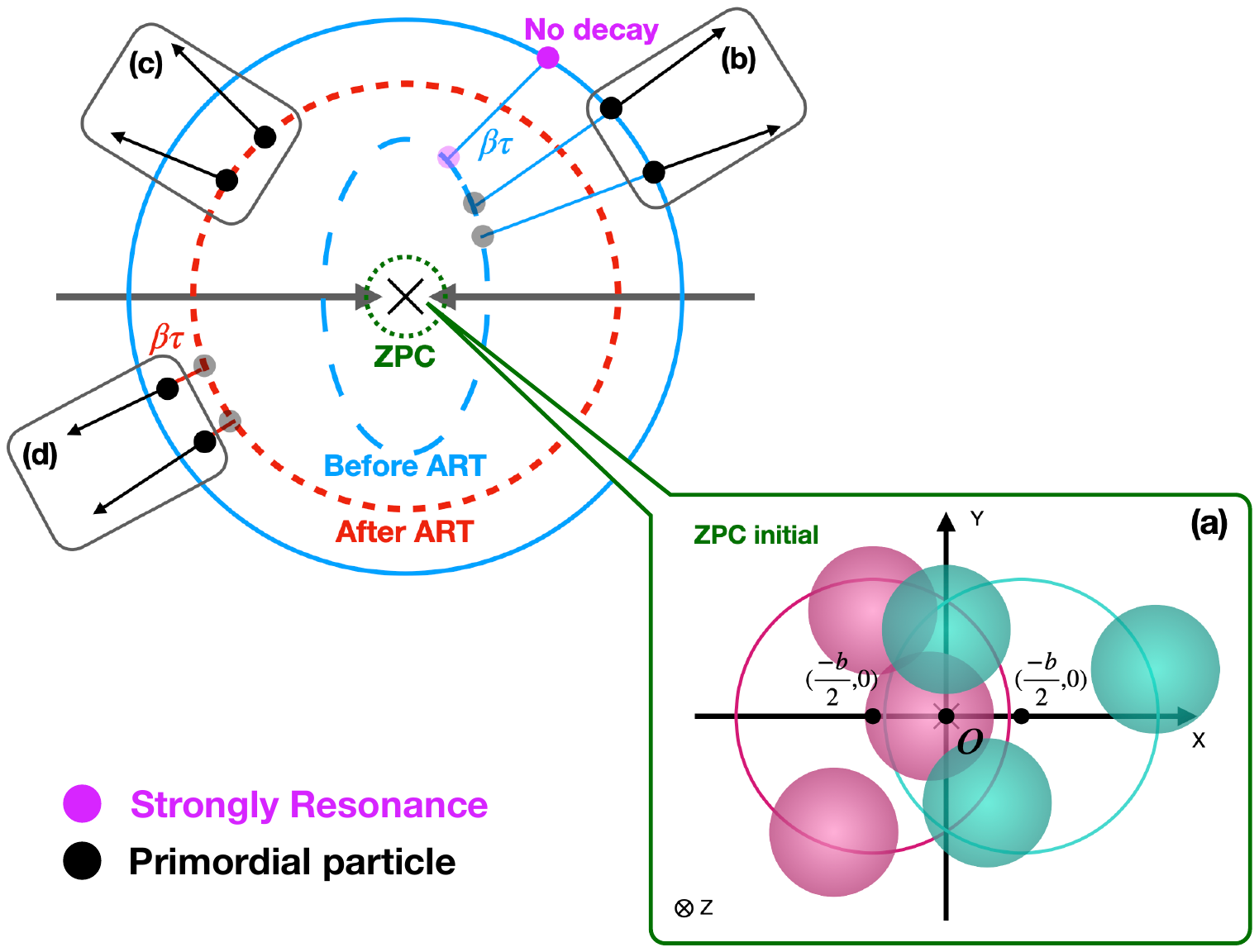}
    \caption{(Color online) A schematic view of the AMPT evolution from the space-time perspective. Panel (a) illustrates the initial partonic distribution. Panels (b) and (d) depict the core source radii for a pair of primordial hadrons, boosted before and after the ART stage, respectively, with an emission time parameter $\tau$ and each particle's velocity $\vec{\beta}$. Panel (c) is the same as (d) but with $\tau=0$. The figure is inspired by~\cite{CECA}.}
    \label{fig:ppCollision}
\end{figure}

The initial partonic distribution during the ZPC stage plays a crucial role in determining the source function. To investigate this effect, three different initial partonic patterns were considered, as demonstrated in the sub-panel (a) of Fig.~\ref{fig:ppCollision}. Partons can be generated from: (1) the overlapping area of quarks (colored disks) inside the protons~\cite{ZL3qAMPT}, mimicking the constituent quark scenario; (2) three fixed black points along the impact parameter direction $\it{b}$, corresponding to the centers of two colliding protons and the center of the impact parameter. In the model's coordinate system, these are located at $x=-b/2$, $x=b/2$ and $x=0$, respectively. This is the intrinsic setting of AMPT though it may not be entirely realistic; (3) the geometrical center of the event ($x=0$), serving as a reference for cross-checking the extreme case. These three settings are labeled as ``3 quarks", ``Normal", and ``Point-like", respectively, in this work.

\begin{figure}
    \centering
    \includegraphics[width=0.5\textwidth]{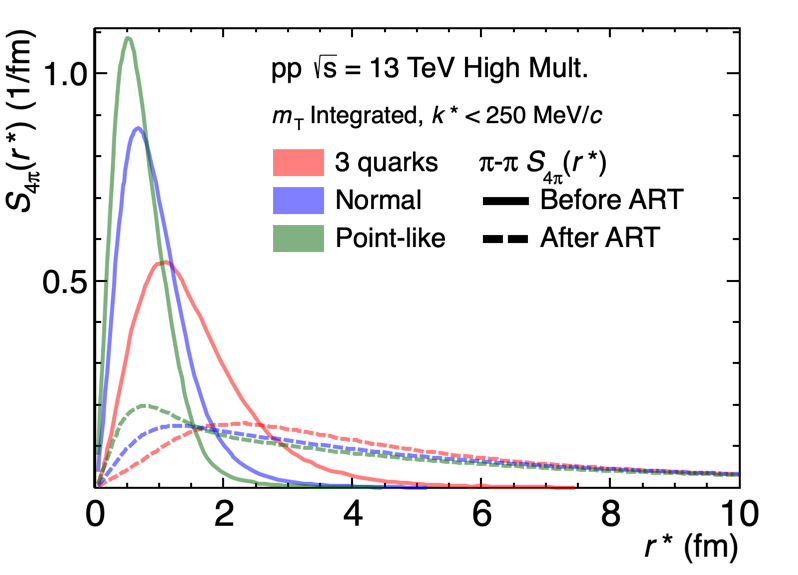}
    \caption{(Color online) The \pipi source functions before and after the ART stage from three initial partonic distributions. See the text for details.}
    \label{fig:ZPCeffectrS}
\end{figure}

In Fig.~\ref{fig:ZPCeffectrS}, the \mt integrated source functions of \pipi pairs at femtoscopy region ($k^{*} < 250$~MeV/$\it{c}$) for three initial configurations are represented. The results at before and after ART stage are represented by solid and dashed lines, respectively. Since the coalescence mechanism from partons to hadrons is identical for any initial partonic configuration, discrepancies in the source function before the ART stage can only arise from differences in the initial partonic distribution. The mean relative distance $\langle r^* \rangle$ of the source function is a convenient variable for comparing different distributions. Qualitative observations of the source function shape show that, before the ART stage, the source aligns with a Gaussian with $R_{\pi}\approx$ 0.65, 0.41, and 0.33 fm and $\langle\it{r^*}\rangle\approx$ 1.58, 0.98, and 0.77 fm for three initial patterns, respectively. In contrast, after the ART stage, the source matches a Cauchy with $R_{\exp}\approx$ 2.22, 2.13, and 1.93 fm and $\langle\it{r^*}\rangle \approx $ 4.15, 3.94, and 3.60 fm. 
As mentioned in Sec.~\ref{sec:method}, the Cauchy source is considered an effective representation of the genuine source, with its exponential tail primarily originating from the resonances, which will be investigated in the following sections. In the ``3 quarks" model, hadrons can only be generated from the overlap region of the binary constituent quark, as shown in sub-panel (a) of Fig.~\ref{fig:ppCollision}. This overlap has the potential to contribute to a more widely dispersed distribution in coordinate space compared to the other two initial distributions, resulting in a broader source function.

\subsection{Fitting source function in different \kT intervals}
\begin{figure}
    \centering
    \includegraphics[width=0.5\textwidth]{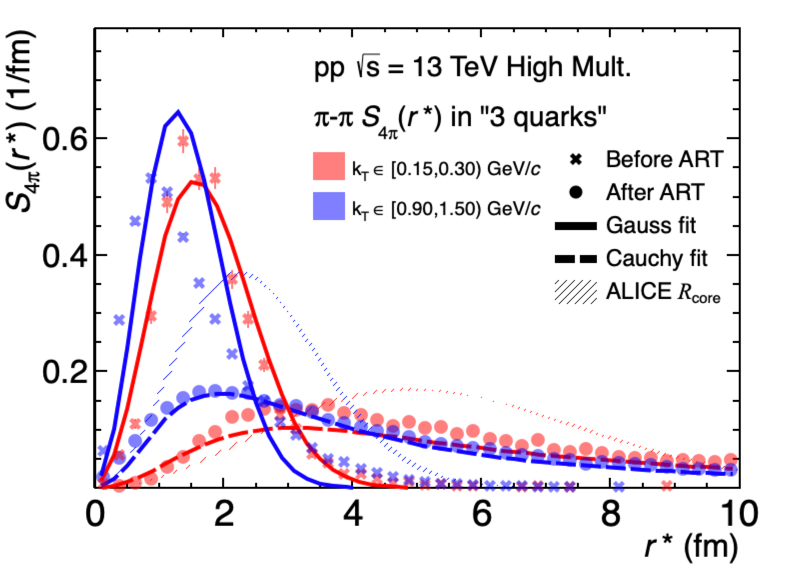}
    \caption{(Color online) The source function in the $k_\text{T} \in [0.15, 0.3)$ and $[0.9, 1.5)$ GeV/$c$ intervals before and after ART stages within the ``3 quarks" AMPT model. Fitting with Gaussian and Cauchy functions are represented by solid and dashed lines, respectively. The shaded bands are the core radii from Ref.~\cite{SourceMaxi}.}
    \label{fig:2kTsource3q}
\end{figure}

The correlation function is commonly divided into different \mt intervals to ensure a consistent number of pairs in the femtoscopic signal region (e.g., $k^{*} < 250$ MeV/\textit{c}). Here the \pipi source function in AMPT is also divided into different \kt ($m_\text{T}$) intervals, following Ref.~\cite{SourceMaxi}, with \kt ranges $0.15\mbox{--}0.3$, $0.3\mbox{--}0.5$, $0.5\mbox{--}0.7$, $0.7\mbox{--}0.9$, and $0.9\mbox{--}1.5$ GeV/\textit{c}. 
The general explanation for the \mt-scaling observed in several different experiments~\cite{alice276,ppSource,SourceMaxi} and simulations~\cite{therm2014,therm2018,CECA} is that a higher $m_\text{T}$ corresponds to particles being generated earlier, leading to smaller source radius. Conversely, as $m_\text{T}$ decreases, low momentum particles are more likely to be produced in more extensive homogeneity region~\cite{LisaReview}, contributing to a larger source. In Fig.~\ref{fig:2kTsource3q}, the source functions for two example $k_\text{T}$ intervals are represented by red and blue lines, with the before and after ART stages shown by solid cross and circle markers, respectively.

Gaussian and Cauchy source functions are used to fit the source distributions before and after ART stage in two $k_\text{T}$ intervals, represented by solid and dashed line, respectively. This fitting yields radii of $R_{\pi} = 0.78$ and $0.63$ fm and $R_{\exp} = 3.09$ and $1.96$ fm, respectively. 
The experimental measurements with corresponding core radii, $R_\text{core} = 2.46 \pm 0.028$ fm and $1.13 \pm 0.015$ fm, are presented by the shaded boxes. According to Fig. 1 of Ref.~\cite{SourceMaxi}, strongly resonances only reduce the height of the peak in the source function and do not change its position. It can be inferred that the core source radius obtained in the ``3 quarks" scenario after the ART stage is smaller than the corresponding core source extracted from experimental data within the given $k_\text{T}$ interval. Given that, as shown in Fig.~\ref{fig:ZPCeffectrS}, the ``3 quarks" scenario already provides the largest size among the three configurations, the ``Normal" and ``Point-like" scenarios are too small to adequately describe the data. The effect of resonances to source function will be explained below.

\subsection{\pipi correlation functions}
\begin{figure}
    \centering
    \includegraphics[width=0.5\textwidth]{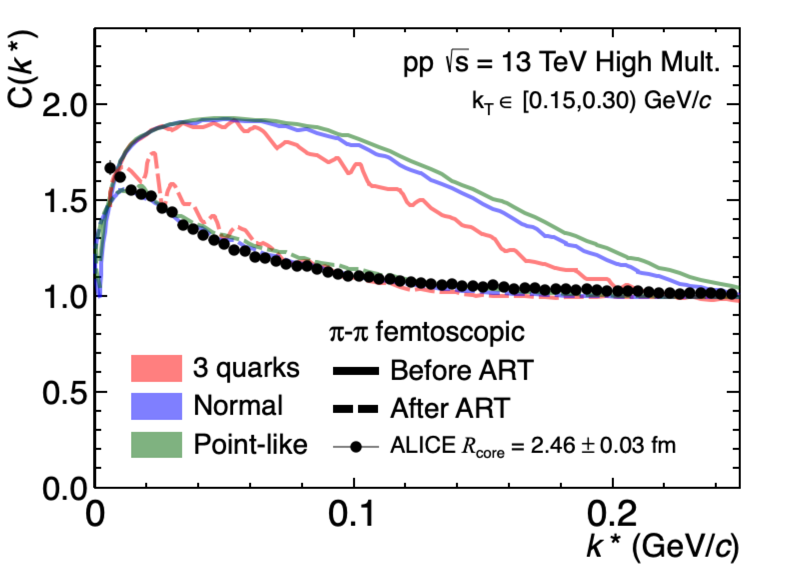}
    \caption{(Color online) The \pipi correlation function in the $k_\text{T}$ interval 0.15–0.30~GeV/$\it{c}$ before and after the ART stages for three initial partonic distributions in AMPT+CATS framework~\cite{Dimi_CATS}.}
    \label{fig:femtoAMPTkt1}
\end{figure}

\begin{figure}
    \centering
    \includegraphics[width=0.5\textwidth]{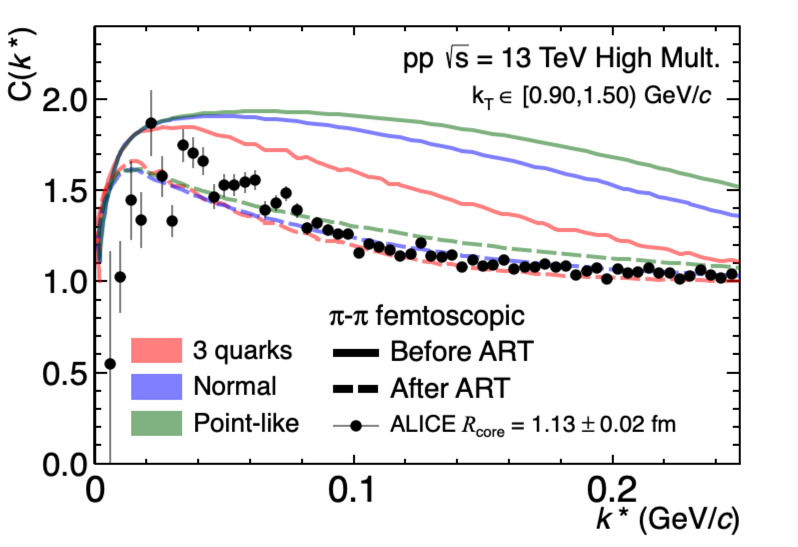}
    \caption{(Color online) Same as Fig.~\ref{fig:femtoAMPTkt1}, but for another $k_\text{T}$ interval 0.90–1.50~GeV/$\it{c}$.}
    \label{fig:femtoAMPTkt5}
\end{figure}

To understand the source's impact on the final correlation function, simulation results are presented for three different initial partonic distributions, before and after the ART stage, using the aforementioned source functions and the accurate two-particle wave function from CATS~\cite{Dimi_CATS}. These results, shown in Fig.~\ref{fig:femtoAMPTkt1} and Fig.~\ref{fig:femtoAMPTkt5}, correspond to two example $k_\text{T}$ intervals.

It can be observed that the correlation functions after the ART stage approximate the experimental data to a certain extent. In contrast, since the source distribution before the ART stage is concentrated in the small \(r^*\) region, the correlation function strength is higher than that observed in the experimental measurements.
Considering only the quantum statistical effects, the correlation function for two identical particles is given in Ref.~\cite{BigBangToSmall} by \( C(k^*) = 1 \pm \exp\left(-k^{*2}R^{2}\right) \), where the \( \pm \) sign corresponds to Bose-Einstein and Fermi-Dirac statistics, respectively.
In this formula, the maximum value of the correlation function \( C(k^*) \) equals 2 at \( k^* = 0 \), and it decreases to 1 as \( k^* \) increases. However, due to the influence of long-range Coulomb interactions, the correlation function is significantly distorted for \( k^* < 50 \) MeV/\(c\). For large \( k^* \), the correlation function approaches unity, with the rate of decrease primarily determined by the shape of the source distribution. If the source is concentrated in a small \( r^* \) region, the correlation function decreases more slowly with \( k^* \). Conversely, if the source is more widely distributed, the larger \( r^* \) regions, where interactions are weaker, contribute more, leading to a larger dilution of the signal. It can also be observed that deviations in \( C(k^*) \) between the three initial partonic conditions occur only at high \( k^* \) before ART and are negligible at low \( k^* \) and after ART. This indicates that, unlike azimuthal observables such as $v_2$ calculated in Refs.~\cite{ZL3qAMPT,Xinlipp13Cumulants,zheng_disentangling_2024}, $C(k^*)$ is less sensitive to the initial geometrical conditions.

\subsection{Impact of parton scattering cross section on the source function}
\begin{figure}
    \centering
    \includegraphics[width=0.49\textwidth]{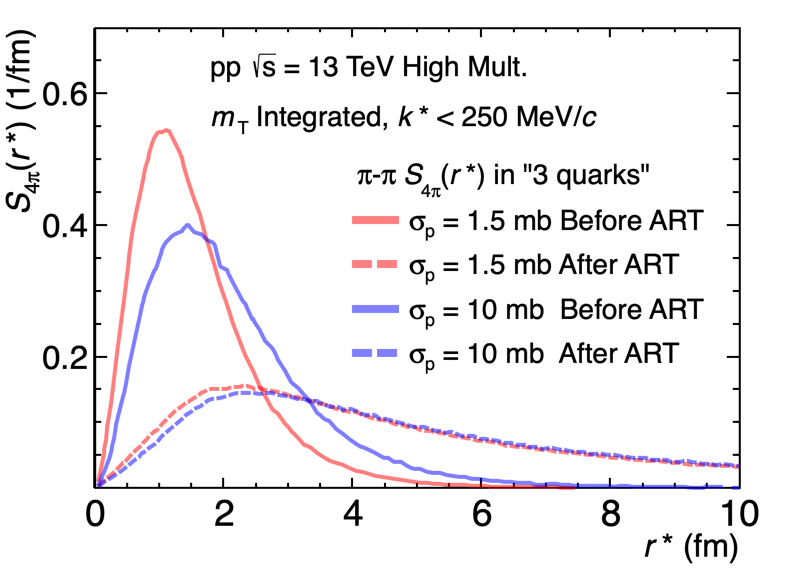}
    \caption{(Color online) The source function for two parton scattering cross-sections, $\sigma_{p}$ = 1.5 mb and 10 mb, before and after the ART stage in the ``3 quarks" scenario.}
    \label{fig:CrossSecEffect}
\end{figure}

Beside the initial position of the parton, the parton scattering cross section $\sigma_{p}$, which reflects the probability of two partons interacting, also affects the source function and the final correlation function, as previously discussed in Ref.~\cite{Li20083DAMPTpiSource}. 
In Fig.~\ref{fig:CrossSecEffect}, the source function for the ``3 quarks" scenario is presented for $\sigma_{p}$ = 1.5 mb and 10 mb. It can be observed that $\sigma_{p}$ significantly affects the source function before the ART stage. The results are similar for the other two initial partonic distributions. As $\sigma_{p}$ increases, the probability of two-parton interactions rises, leading to a more dispersed parton and hadron distribution, and consequently, a wider source function. However, the results after ART are almost unaffected by $\sigma_{p}$, indicating that the hadronic process plays a decisive role. Most initial effects are smeared or masked by hadronic scattering and resonance decay, which will be discussed in the following section.

\subsection{Impact of resonance and hadronic scattering on the source function}
\begin{figure}
    \centering
    \includegraphics[width=0.5\textwidth]{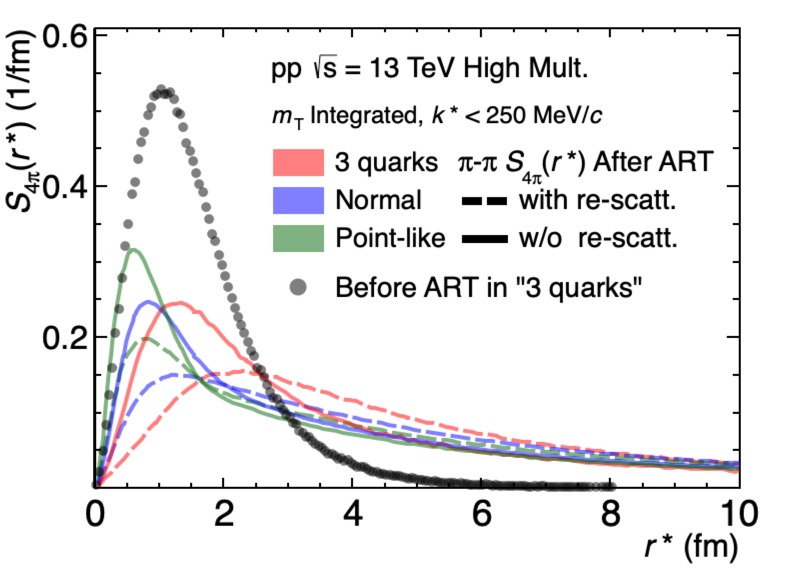}
    \caption{(Color online) The source function, with and without hadronic rescattering, after the ART stage for three initial partonic distributions.}
    \label{fig:NTMax3effect}
\end{figure}

The hadronic interaction in AMPT, ART, is dominated by two mechanisms: short-lived strongly-decaying resonances and hadronic rescattering including both elastic and inelastic processes. 
In Fig.~\ref{fig:NTMax3effect}, it can be seen that, compared to the case where rescattering is turned off, the \rstar distribution is wider when rescattering is on for all three initial conditions. This aligns with the expectation that the generated hadrons undergo adequate hadronic scatterings, causing the entire system to expand outward. The long tail persists even when rescattering is off, suggesting the contribution from the resonance decay.

As mentioned in Sec.~\ref{sec:method}, the source function has two main components: the primordial particles produced in collisions, which are well described by a Gaussian distribution with width \( R_\text{core} \) (the core part), and a non-Gaussian contribution, represented by an exponential tail, mainly arising from short-lived resonances.
Previous study shows that the Statistical Hadronization Model (SHM)~\cite{THERMINATOR2} combined with EPOS~\cite{EPOSmodel}, can well reproduce the tail part measured in ALICE~\cite{ppSource,SourceMaxi}. Using this approach, the core of the source was separated from the ``resonance halo". 

\begin{figure}
    \centering
    \includegraphics[width=0.5\textwidth]{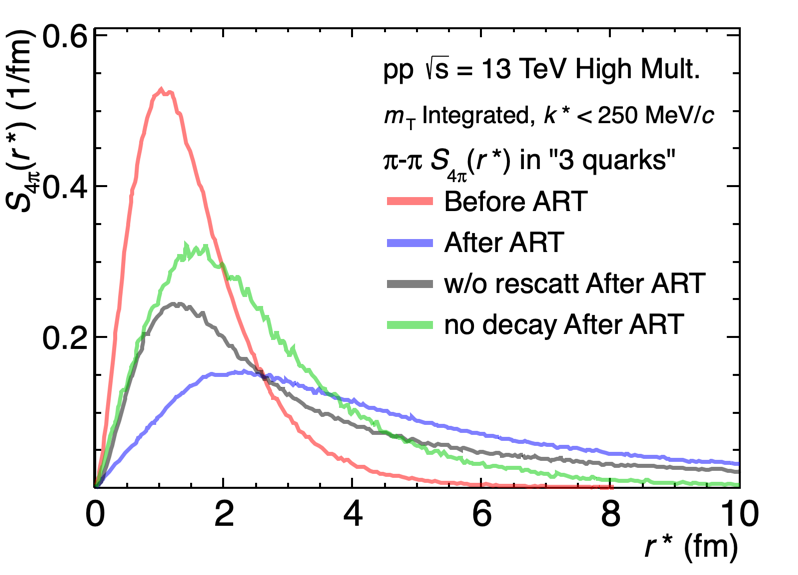}
    \caption{(Color online) The source functions at four scenarios in the ``3 quarks" AMPT. The black and green lines represent the results after ART, with the black line indicating no hadronic rescattering but including resonance decay, and the green line indicating no resonance decay but including hadronic rescattering. See the text for details.}
    \label{fig:3qdetail}
\end{figure}

In the SHM calculation, only the decay products of short-lived resonances that contribute at least 1\% are considered. As shown in Tab.~\ref{Tab:StrongDecayTable}, 28\% of the charged pions are primordial, while 72\% originate from resonances. In AMPT, however, serval decay channels are not included, leading to a different fraction compared to the SHM calculation. There are ongoing efforts to incorporate the production and annihilation channels in the ART stage~\cite{d_productionInAMPT}, and a more thorough description of resonances remains to be explored.
Despite the differences in resonance components and fractions, the qualitative impact of FSIs on the source is shown in Fig.~\ref{fig:3qdetail}.
Four scenarios are investigated: before ART, after ART, after ART without the rescattering process, and after ART without resonance decay but with hadronic rescattering. The total ART contribution (blue) can be decomposed into the resonance (black) and hadronic rescattering (green) components. In the case where no resonances contribute to the pions (green), the tail of the relative source function is significantly shorter compared to the gray one, which extends up to 30 fm.

\begin{table}
\centering
\caption{List of resonances contributing to the yield of $\pi$ in HM pp collisions at $\sqrt{s}=13$~TeV. The left column is from Ref.~\cite{SourceMaxi}, calculated using the THERMAL-FIST package, while the right column shows the AMPT results without kinematic cuts.}

\begin{tabular}{c|c|c}

\hline
 & SHM Fraction(\%)& ``3 quarks" Fraction(\%)\\
\hline
primordial & 28.0 & 46.3\\
\hline
strong resonances &72.0 & 53.7 \\
\hline
\hline
Resonances \\
\hline
$\rho(770)^{0}$ & 9.0 & 6.8 \\
\hline
$\rho(770)^{+}$ & 8.7 & 13.9\\
\hline
$\omega(782)$ & 7.7 & 6.2\\
\hline
$K^{*}(892)^{+}$ & 2.3 & 4.3 \\
\hline
$\bar{K}^{*}(892)^{0}$ &2.6 & 4.2 \\
\hline
$b_{1}(1235)^{0}$ & 1.9 & -\\
\hline
$a_{2}(1320)^{+}$ & 1.5 & -  \\
\hline
$\eta$ & 1.5 & 19.9  \\
\hline
$a_{1}(1260)^{+}$ & 1.4 & - \\
\hline
$f_{2}(1270)$ & 1.4 & -  \\
\hline
$a_{0}(980)^{+}$ & 1.4 & - \\
\hline
$h_{1}(1170)$ & 1.2 & - \\
\hline
\end{tabular}

\label{Tab:StrongDecayTable}
\end{table}

\subsection{Final source function and \mt-scaling}
\begin{figure}
    \centering
    \includegraphics[width=0.5\textwidth]{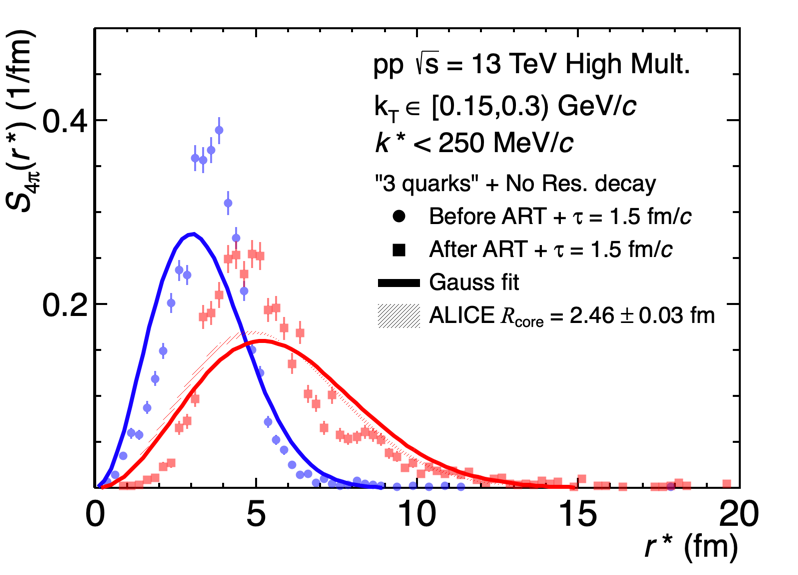}
    \caption{(Color online) The fitting results for the \rstar distribution in the $\kt\in[0.15,0.3]$ \UGeVc. The blue and red markers represent results before and after the ART stage, respectively, with $\tau$ = 1.5 fm/$c$. The shadow band represents the source distribution extracted by ALICE~\cite{SourceMaxi}.}
    \label{fig:detailfit}
\end{figure}

In principle, the standard method for subtracting the resonance contribution from the total source function and extracting $R_\text{core}$ is to follow Eq. (4) in Ref.~\cite{ppSource}, which employs Gaussian fitting of $S_\text{total}(r^*)$ to decompose the primordial and resonance components.
However, whether this approach is applicable within the AMPT framework remains to be examined. 
In this study, an alternative method is employed. The schematic representation of the space-time dynamics is shown in Fig~\ref{fig:ppCollision}. Collisions with a given initial distribution (panel a) first proceed through the ZPC stage (green dashed circle). After the coalescence process, hadrons are formed (blue dashed circle), which represents the stage before ART. To understand the core source function, the default resonance decays in ART are fully turned off so their contributions are excluded, matching the original definition of the core source and the scenario described in Ref.~\cite{CECA}. An emission time parameter $\tau$ is then introduced. The generated hadrons are forced to travel along their original momentum directions for $\tau$ fm/$c$ without any hardonic interactions, resulting in the boost of the core source radii by $\vec{\beta}\tau$ fm ($\vec{\beta}$ is particle's velocity) in spatial coordinates (panel b). For comparison, the default ART process including hadronic interactions (red dashed circle) is also studied with a possible boost, where $\tau$ can be zero (panel c) or non-zero (panel d).
Figure~\ref{fig:detailfit} shows the \rstar distribution (dots) and the fitting results using Gaussians (lines) in the \kt interval of 0.15--0.3 GeV/$c$. The emission time parameter $\tau=1.5$ fm/$\it{c}$ is computed using the weighted abundances and lifetimes of the resonances considered in Ref.~\cite{SourceMaxi}. It is observed that the average radius after ART (red) is lager than that before ART (blue) as expected. The fittings work approximately despite minor inaccuracies. It is worthy noting that the result after ART is highly compatible with the ALICE result~\cite{SourceMaxi}, indicating the validity of the model.

\begin{figure}
    \centering
    \includegraphics[width=0.5\textwidth]{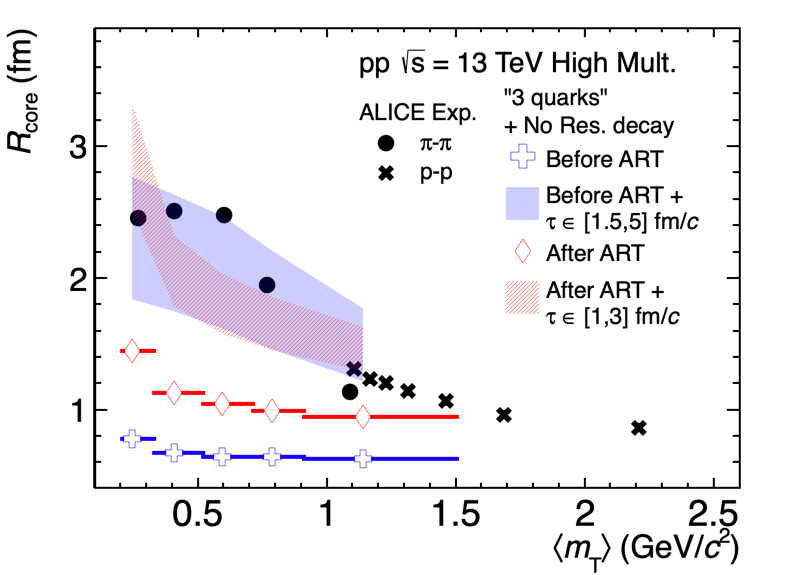}
    \caption{(Color online) The \mt-scaling behavior of $R_\text{core}$ in AMPT is shown for four emission scenarios (see the text for details), and is compared with the ALICE results (solid markers).}
    \label{fig:FinalmTscaling}
\end{figure}

Based the on the fitting results, the \rcore values in each \kt interval are extracted for four different scenarios: (\rmnum{1}) before ART without further boost, (\rmnum{2}) before ART with a boost of $\tau\in[1.5,5]$ fm/$\it{c}$, (\rmnum{3}) after ART stage without boost, and (\rmnum{4}) after ART with a boost of $\tau\in[1,3]$ fm/$\it{c}$. In (\rmnum{2}), the upper limit originates from the general assumption that resonances with $\textit{c}\tau_\text{res} > 5$ fm are long-lived, while in (\rmnum{4}), the $\tau$ values are deliberately reduced to roughly match the results of (\rmnum{2}). Figure ~\ref{fig:FinalmTscaling} shows the \mt-scaling behavior of \rcore. It can be seen that all four cases are in line with the expectation that the source radii decrease as \mt increases, roughly following the power-law relationship~\cite{SourceMaxi}: $R_\text{core} = \it{a} + \it{b}\cdot\langle m_\text{T}\rangle^{\it{c}}$. This can be understood in terms of the collectivity generated since the partonic stage. Compared to the ALICE measurements (solid dots), the original AMPT sources (\rmnum{1}) and (\rmnum{3}), without additional boosting, are systematically smaller, while the modified cases of (\rmnum{2}) and (\rmnum{4}) are in good agreement with data in the low \mt ranges. Note that ALICE results also exhibit a plateau at $\mt < 0.5$~GeV/$\it{c}^\mathrm{2}$, which can be interpreted as the limitation of the system size in pp collisions. However, this feature is not observed in AMPT, which instead follows the power-law increasing tread. Generally, the AMPT provides a good environment to reveal the mechanisms behind system size and \mt-scaling, but further investigation is also required to understand the detailed behaviors.

\section{summary}\label{sec:summary}
This paper presents an detailed investigation of the pion emission source in high multiplicity pp collisions at $\sqrt{s}=13$~TeV using the AMPT model with different initial partonic distributions including one with the sub-nucleon structure. 
Source functions and corresponding correlation functions are calculated, with the latter obtained using the precise two-particle FSI from the CATS framework. Results show that the initial partonic distribution influences the source size, with the ``3 quarks" mode generating a relatively larger radius due to its larger initial spatial distribution. The $k_\text{T}$ dependence is observed and the partonic scattering cross section also plays a role. However, when it comes to the correlation function, the initial effects are largely smeared by FSIs. 
To understand the final state dynamics, two main components, the resonance decay and hadronic scatterings, are carefully studied. The long tail in the source function is attributed to the resonances. After tuning off the decay process, the core source radii $R_\text{core}$ can be extracted. A clear \mt-scaling behavior is observed, and with an appropriate emission time, $R_\text{core}$ in AMPT can well reproduce the ALICE measurements, providing new insights into the space-time characteristics, particle generation mechanisms in pp collisions, and potential improvements to the AMPT model.

For future studies, the resonance decay channels in ART should be updated. The relationship between radial (anisotropic) flow and the source function needs to be quantified. Studies on other particle species (e.g. \pP and \Kp pairs) and the source function in multiple dimensions would also be valuable for a better understanding of the experimental measurements.

\section*{ACKNOWLEDGMENTS}
We are grateful to the femTUM group, led by Laura Fabbietti, for enlightening discussions and suggestions, in particular, Maximilian Korwieser and Dimitar Mihaylov.

\bibliographystyle{utphys}
\bibliography{bibliography}

\providecommand{\href}[2]{#2}\begingroup\raggedright\begin{thebibliography}{10}

\bibitem{LisaReview}
M.~A. Lisa, S.~Pratt, R.~Soltz, and U.~Wiedemann, ``{Femtoscopy in relativistic heavy ion collisions}'', \href{http://dx.doi.org/10.1146/annurev.nucl.55.090704.151533}{{\em Ann. Rev. Nucl. Part. Sci.} {\bfseries 55} (2005) 357--402}, \href{http://arxiv.org/abs/nucl-ex/0505014}{{\ttfamily arXiv:nucl-ex/0505014}}.

\bibitem{LauraFreview}
L.~Fabbietti, V.~Mantovani~Sarti, and O.~Vazquez~Doce, ``{Study of the Strong Interaction Among Hadrons with Correlations at the LHC}'', \href{http://dx.doi.org/10.1146/annurev-nucl-102419-034438}{{\em Ann. Rev. Nucl. Part. Sci.} {\bfseries 71} (2021) 377--402}, \href{http://arxiv.org/abs/2012.09806}{{\ttfamily arXiv:2012.09806 [nucl-ex]}}.

\bibitem{QMgriffiths1984}
R.~B. Griffiths, ``{Consistent histories and the interpretation of quantum mechanics}'', {\em Journal of Statistical Physics} {\bfseries 36} (1984) 219--272.

\bibitem{Lednicky:1981su}
R.~Lednicky and V.~L. Lyuboshits, ``{Final State Interaction Effect on Pairing Correlations Between Particles with Small Relative Momenta}'', {\em Sov. J. Nucl. Phys.} {\bfseries 35} (1982) 770. [Yad. Fiz.35,1316(1981)].

\bibitem{LednickFinalSize}
R.~Lednick{\'y}, ``{Finite-size effect on two-particle production in continuous and discrete spectrum}'', \href{http://dx.doi.org/10.1134/S1063779609030034}{{\em Physics of Particles and Nuclei} {\bfseries 40} (2009) 307--352}. \url{https://doi.org/10.1134/S1063779609030034}.

\bibitem{ppFemtoSTAR}
L.~Adamczyk {\em et~al.}, ``{Measurement of interaction between antiprotons}'', \href{http://dx.doi.org/10.1038/nature15724}{{\em Nature} {\bfseries 527} (2015) 345--348}. \url{https://doi.org/10.1038/nature15724}.

\bibitem{alice276}
{\bfseries ALICE} Collaboration, J.~Adam {\em et~al.}, ``{One-dimensional pion, kaon, and proton femtoscopy in Pb--Pb collisions at $\sqrt{s_{\rm{NN}}}$ = 2.76 TeV}'', \href{http://dx.doi.org/10.1103/PhysRevC.92.054908}{{\em Phys. Rev. C} {\bfseries 92} (Nov, 2015) 054908}.

\bibitem{ks0kPbALICE}
S.~Acharya {\em et~al.}, ``{Measuring $K_{s}^{0}K^{\pm}$ interactions using Pb--Pb collisions at $\sqrt{s_{\rm{NN}}}$ = 2.76 TeV}'', \href{http://dx.doi.org/https://doi.org/10.1016/j.physletb.2017.09.009}{{\em Physics Letters B} {\bfseries 774} (2017) 64--77}. \url{https://www.sciencedirect.com/science/article/pii/S0370269317307074}.

\bibitem{kpPbALICE}
S.~Acharya, , {\em et~al.}, ``{Kaon–proton strong interaction at low relative momentum via femtoscopy in Pb--Pb collisions at the LHC}'', \href{http://dx.doi.org/https://doi.org/10.1016/j.physletb.2021.136708}{{\em Physics Letters B} {\bfseries 822} (2021) 136708}. \url{https://www.sciencedirect.com/science/article/pii/S0370269321006481}.

\bibitem{Pioninterf1979}
M.~Gyulassy, S.~Kauffmann, and L.~W. Wilson, ``{Pion interferometry of nuclear collisions. I. Theory}'', {\em Physical Review C} {\bfseries 20} (1979) 2267.

\bibitem{Pioninterf1991}
A.~Capella, A.~Krzywicki, and E.~M. Levin, ``{Pion interferometry and intermittency in heavy-ion collisions}'', \href{http://dx.doi.org/10.1103/PhysRevD.44.704}{{\em Phys. Rev. D} {\bfseries 44} (Aug, 1991) 704--716}. \url{https://link.aps.org/doi/10.1103/PhysRevD.44.704}.

\bibitem{BEC2010_pp900}
K.~Aamodt {\em et~al.}, ``{Two-pion Bose-Einstein correlations in pp collisions at $\sqrt{s}$ = 900 GeV}'', \href{http://dx.doi.org/10.1103/physrevd.82.052001}{{\em Physical Review D} {\bfseries 82} (2010) }. \url{http://dx.doi.org/10.1103/PhysRevD.82.052001}.

\bibitem{BEC2011_PbPb276}
K.~Aamodt {\em et~al.}, ``{Two-pion Bose--Einstein correlations in central Pb--Pb collisions at $\sqrt{s_{\rm{NN}}}$ = 2.76 TeV}'', {\em Physics Letters B} {\bfseries 696} (2011) 328--337.

\bibitem{BEC1988}
M.~G. Bowler, ``{Extended sources, final state interactions and Bose-Einstein correlations}'', \href{http://dx.doi.org/10.1007/BF01560395}{{\em Zeitschrift f{\"u}r Physik C Particles and Fields} {\bfseries 39} (1988) 81--88}. \url{https://doi.org/10.1007/BF01560395}.

\bibitem{pipiScatter}
G.~Colangelo, J.~Gasser, and H.~Leutwyler, ``{$\pi\pi$ scattering}'', \href{http://dx.doi.org/10.1016/s0550-3213(01)00147-x}{{\em Nuclear Physics B} {\bfseries 603} (June, 2001) 125–179}. \url{http://dx.doi.org/10.1016/S0550-3213(01)00147-X}.

\bibitem{pipiEffectRange}
S.~K. Adhikari and J.~{R.A. Torreão}, ``{Effective range expansion for the pion-pion system}'', \href{http://dx.doi.org/https://doi.org/10.1016/0370-2693(83)90992-9}{{\em Physics Letters B} {\bfseries 123} (1983) 452--454}. \url{https://www.sciencedirect.com/science/article/pii/0370269383909929}.

\bibitem{SourceMaxi}
{\bfseries ALICE} Collaboration, ``{Common femtoscopic hadron-emission source in pp collisions at the LHC}'', 2023.
\newblock \url{https://arxiv.org/abs/2311.14527}.

\bibitem{ppSource}
{\bfseries ALICE} Collaboration, S.~Acharya {\em et~al.}, ``{Search for a common baryon source in high-multiplicity pp collisions at the LHC}'', \href{http://dx.doi.org/10.1016/j.physletb.2020.135849}{{\em Phys. Lett. B} {\bfseries 811} (2020) 135849}, \href{http://arxiv.org/abs/2004.08018}{{\ttfamily arXiv:2004.08018 [nucl-ex]}}.

\bibitem{therm2014}
A.~Kisiel, M.~Gałażyn, and P.~Bożek, ``{Pion, kaon, and proton femtoscopy in Pb--Pb collisions at $\sqrt{s_{\rm NN}}=2.76$ TeV modeled in (3+1)D hydrodynamics}'', \href{http://dx.doi.org/10.1103/PhysRevC.90.064914}{{\em Phys. Rev. C} {\bfseries 90} (Dec, 2014) 064914}.

\bibitem{therm2018}
A.~Kisiel, ``{Pion-kaon femtoscopy in Pb–Pb collisions at $\sqrt{s_{\rm NN}}=2.76$ TeV modeled in (3+1)D hydrodynamics coupled to Therminator 2 and the effect of delayed kaon emission}'', \href{http://dx.doi.org/10.1103/PhysRevC.98.044909}{{\em Phys. Rev. C} {\bfseries 98} (Oct, 2018) 044909}.

\bibitem{yu2019}
V.~M. Shapoval and {\relax Yu}.~M. Sinyukov, ``{Bulk observables in Pb--Pb collisions at $\sqrt{{s}_{\rm NN}} = 5.02$ TeV at the CERN Large Hadron Collider within the integrated hydrokinetic model}'', \href{http://dx.doi.org/10.1103/PhysRevC.100.044905}{{\em Phys. Rev. C} {\bfseries 100} (Oct, 2019) 044905}.

\bibitem{Shapoval2014}
V.~Shapoval, P.~Braun-Munzinger, I.~Karpenko, and Y.~Sinyukov, ``{Femtoscopy correlations of kaons in Pb--Pb collisions at LHC within hydrokinetic model}'', \href{http://dx.doi.org/https://doi.org/10.1016/j.nuclphysa.2014.05.003}{{\em Nuclear Physics A} {\bfseries 929} (2014) 1--8}. \url{https://www.sciencedirect.com/science/article/pii/S0375947414001183}.

\bibitem{THERMINATOR2}
M.~Chojnacki, A.~Kisiel, W.~Florkowski, and W.~Broniowski, ``{THERMINATOR 2: THERMal heavy IoN generATOR 2}'', \href{http://dx.doi.org/https://doi.org/10.1016/j.cpc.2011.11.018}{{\em Computer Physics Communications} {\bfseries 183} (2012) 746--773}. \url{https://www.sciencedirect.com/science/article/pii/S0010465511003808}.

\bibitem{ppflow1}
{\bfseries ALICE} Collaboration, S.~Acharya {\em et~al.}, ``Investigations of anisotropic flow using multiparticle azimuthal correlations in $pp$, {$p\text{\ensuremath{--}}\mathrm{Pb}$, Xe--Xe, and Pb--Pb} collisions at the {LHC}'', \href{http://dx.doi.org/10.1103/PhysRevLett.123.142301}{{\em Phys. Rev. Lett.} {\bfseries 123} (Oct, 2019) 142301}. \url{https://link.aps.org/doi/10.1103/PhysRevLett.123.142301}.

\bibitem{ppflow2}
{\bfseries ALICE} Collaboration, S.~Acharya {\em et~al.}, ``Emergence of long-range angular correlations in low-multiplicity proton-proton collisions'', \href{http://dx.doi.org/10.1103/PhysRevLett.132.172302}{{\em Phys. Rev. Lett.} {\bfseries 132} (Apr, 2024) 172302}. \url{https://link.aps.org/doi/10.1103/PhysRevLett.132.172302}.

\bibitem{ppflow3}
{\bfseries ALICE} Collaboration, S.~Acharya {\em et~al.}, ``Observation of partonic flow in proton-proton and proton-nucleus collisions'', 2024.
\newblock \url{https://arxiv.org/abs/2411.09323}.

\bibitem{ppflow4}
{W. Wu (for the ALICE Collaboration)}, ``Probing partonic collectivity in pp and {p--Pb} collisions with {ALICE}'', 2023.
\newblock \url{https://indico.cern.ch/event/1043736/contributions/5363771/}. talk given at IS2023.

\bibitem{EPOSmodel}
T.~Pierog, I.~Karpenko, J.~M. Katzy, E.~Yatsenko, and K.~Werner, ``{EPOS LHC: Test of collective hadronization with data measured at the CERN Large Hadron Collider}'', \href{http://dx.doi.org/10.1103/PhysRevC.92.034906}{{\em Phys. Rev. C} {\bfseries 92} (Sep, 2015) 034906}. \url{https://link.aps.org/doi/10.1103/PhysRevC.92.034906}.

\bibitem{UrQMD1}
M.~Bleicher, {\em et~al.}, ``{Relativistic hadron-hadron collisions in the ultra-relativistic quantum molecular dynamics model}'', \href{http://dx.doi.org/10.1088/0954-3899/25/9/308}{{\em Journal of Physics G: Nuclear and Particle Physics} {\bfseries 25} (Sept., 1999) 1859–1896}. \url{http://dx.doi.org/10.1088/0954-3899/25/9/308}.

\bibitem{UrQMD2}
S.~Bass, ``{Microscopic models for ultrarelativistic heavy ion collisions}'', \href{http://dx.doi.org/10.1016/s0146-6410(98)00058-1}{{\em Progress in Particle and Nuclear Physics} {\bfseries 41} (1998) 255–369}. \url{http://dx.doi.org/10.1016/S0146-6410(98)00058-1}.

\bibitem{li_effects_2022}
P.~Li, J.~Steinheimer, T.~Reichert, A.~Kittiratpattana, M.~Bleicher, and Q.~Li, ``Effects of a phase transition on two-pion interferometry in heavy ion collisions at \$\${\textbackslash}sqrt \{\{s\_\{\{{\textbackslash}rm\{{nn}\}\}\}\}\} = 2.4 - 7.7{\textbackslash},{\textbackslash},\{{\textbackslash}rm\{{GeV}\}\}\$\$'', \href{http://dx.doi.org/10.1007/s11433-022-2041-8}{{\em Science China Physics, Mechanics \& Astronomy} {\bfseries 66} (Dec., 2022) 232011}. \url{https://doi.org/10.1007/s11433-022-2041-8}.

\bibitem{li_transport_2022}
P.~Li, Y.~Wang, Q.~Li, and H.~Zhang, ``Transport model analysis of the pion interferometry in {Au}+{Au} collisions at {Ebeam}=1.23 {GeV}/nucleon'', \href{http://dx.doi.org/10.1007/s11433-022-2026-5}{{\em Science China Physics, Mechanics \& Astronomy} {\bfseries 66} (Dec., 2022) 222011}. \url{https://doi.org/10.1007/s11433-022-2026-5}.

\bibitem{fang_azimuthal-sensitive_2022}
L.-M. Fang, Y.-G. Ma, and S.~Zhang, ``Azimuthal-sensitive three-dimensional {HBT} radius in {Au}–{Au} collisions at \$\${E}\_\{beam\} = 1.{23A}\$\${GeV} by the {IQMD} model'', \href{http://dx.doi.org/10.1140/epja/s10050-022-00722-w}{{\em The European Physical Journal A} {\bfseries 58} (Apr., 2022) 81}. \url{https://doi.org/10.1140/epja/s10050-022-00722-w}.

\bibitem{fang_simulation_2023}
L.-M. Fang, Y.-G. Ma, and S.~Zhang, ``Simulation of collective flow of protons and deuterons in \${\textbackslash}mathrm\{{Au}\}+{\textbackslash}mathrm\{{Au}\}\$ collisions at \$\{{E}\}\_\{{\textbackslash}text\{beam\}\}=1.{23A} {\textbackslash}mathrm\{{GeV}\}\$ with the isospin-dependent quantum molecular dynamics model'', \href{http://dx.doi.org/10.1103/PhysRevC.107.044904}{{\em Physical Review C} {\bfseries 107} (Apr., 2023) 044904}. \url{https://link.aps.org/doi/10.1103/PhysRevC.107.044904}.

\bibitem{li_probing_2021}
L.-Y. Li, P.~Ru, and Y.~Hu, ``Probing granular inhomogeneity of a particle-emitting source by imaging two-pion {Bose}–{Einstein} correlations'', \href{http://dx.doi.org/10.1007/s41365-021-00853-7}{{\em Nuclear Science and Techniques} {\bfseries 32} (Feb., 2021) 19}. \url{https://doi.org/10.1007/s41365-021-00853-7}.

\bibitem{CRAB}
S.~Pratt, {\em et~al.}, ``{Testing transport theories with correlation measurements}'', \href{http://dx.doi.org/https://doi.org/10.1016/0375-9474(94)90614-9}{{\em Nuclear Physics A} {\bfseries 566} (1994) 103--114}. \url{https://www.sciencedirect.com/science/article/pii/0375947494906149}.

\bibitem{Li2002AMPTpiSource}
Z.-W. Lin, C.~M. Ko, and S.~Pal, ``{Partonic Effects on Pion Interferometry at the Relativistic Heavy-Ion Collider}'', \href{http://dx.doi.org/10.1103/physrevlett.89.152301}{{\em Physical Review Letters} {\bfseries 89} (Sept., 2002) }. \url{http://dx.doi.org/10.1103/PhysRevLett.89.152301}.

\bibitem{Li20083DAMPTpiSource}
Z.-W. Lin, ``{3D pion source functions from the AMPT model}'', \href{http://dx.doi.org/10.1088/0954-3899/35/10/104138}{{\em Journal of Physics G: Nuclear and Particle Physics} {\bfseries 35} (Sep, 2008) 104138}. \url{https://dx.doi.org/10.1088/0954-3899/35/10/104138}.

\bibitem{piSourceNICA}
A.~Ayala, S.~Bernal-Langarica, I.~Dominguez, I.~Maldonado, and M.~E. Tejeda-Yeomans, ``{Collision energy dependence of source sizes for primary and secondary pions at NICA energies}'', 2024.
\newblock \url{https://arxiv.org/abs/2401.00619}.

\bibitem{piSourceEPOS}
M.~Stefaniak and D.~Kincses, \href{http://dx.doi.org/10.1117/12.2580570}{``{Investigating the pion source function in heavy-ion collisions with the EPOS model}'',} in {\em Photonics Applications in Astronomy, Communications, Industry, and High Energy Physics Experiments 2020}, R.~S. Romaniuk and M.~Linczuk, eds., vol.~36, p.~37.
\newblock SPIE, Oct., 2020.
\newblock \url{http://dx.doi.org/10.1117/12.2580570}.

\bibitem{UrQMDpiSource}
E.~Khyzhniak, V.~Semenova, N.~Ermakov, and G.~Nigmatkulov, ``{Estimation of pion-emitting source in symmetric and asymmetric collisions using the UrQMD model}'', in {\em EPJ Web of Conferences}, vol.~204, p.~03017, EDP Sciences.
\newblock 2019.

\bibitem{pipiUrQMDHIC}
N.~Ermakov and G.~Nigmatkulov, ``{Modeling of two-particle femtoscopic correlations at top RHIC energy}'', \href{http://dx.doi.org/10.1088/1742-6596/798/1/012055}{{\em Journal of Physics: Conference Series} {\bfseries 798} (Jan., 2017) 012055}. \url{http://dx.doi.org/10.1088/1742-6596/798/1/012055}.

\bibitem{CECA}
D.~Mihaylov and J.~Gonz{\'a}lez~Gonz{\'a}lez, ``{Novel model for particle emission in small collision systems}'', \href{http://dx.doi.org/10.1140/epjc/s10052-023-11774-7}{{\em The European Physical Journal C} {\bfseries 83} (2023) 590}. \url{https://doi.org/10.1140/epjc/s10052-023-11774-7}.

\bibitem{ZL3qAMPT}
L.~Zheng, G.-H. Zhang, Y.-F. Liu, Z.-W. Lin, Q.-Y. Shou, and Z.-B. Yin, ``{Investigating high energy proton proton collisions with a multi-phase transport model approach based on PYTHIA8 initial conditions}'', \href{http://dx.doi.org/10.1140/epjc/s10052-021-09527-5}{{\em The European Physical Journal C} {\bfseries 81} (2021) 755}. \url{https://doi.org/10.1140/epjc/s10052-021-09527-5}.

\bibitem{Xinlipp13Cumulants}
X.-L. Zhao, Z.-W. Lin, L.~Zheng, and G.-L. Ma, ``{A transport model study of multiparticle cumulants in pp collisions at 13 TeV}'', \href{http://dx.doi.org/https://doi.org/10.1016/j.physletb.2023.137799}{{\em Physics Letters B} {\bfseries 839} (2023) 137799}. \url{https://www.sciencedirect.com/science/article/pii/S0370269323001338}.

\bibitem{zheng_disentangling_2024}
L.~Zheng, L.~Liu, Z.-W. Lin, Q.-Y. Shou, and Z.-B. Yin, ``Disentangling the development of collective flow in high energy proton proton collisions with a multiphase transport model'', \href{http://dx.doi.org/10.1140/epjc/s10052-024-13378-1}{{\em The European Physical Journal C} {\bfseries 84} (Oct., 2024) 1029}. \url{https://doi.org/10.1140/epjc/s10052-024-13378-1}.

\bibitem{AMPTorigin}
Z.-W. Lin, C.~M. Ko, B.-A. Li, B.~Zhang, and S.~Pal, ``{Multiphase transport model for relativistic heavy ion collisions}'', \href{http://dx.doi.org/10.1103/PhysRevC.72.064901}{{\em Phys. Rev. C} {\bfseries 72} (Dec, 2005) 064901}. \url{https://link.aps.org/doi/10.1103/PhysRevC.72.064901}.

\bibitem{lin_further_2021}
Z.-W. Lin and L.~Zheng, ``Further developments of a multi-phase transport model for relativistic nuclear collisions'', \href{http://dx.doi.org/10.1007/s41365-021-00944-5}{{\em Nuclear Science and Techniques} {\bfseries 32} (Oct., 2021) 113}. \url{https://link.springer.com/10.1007/s41365-021-00944-5}.

\bibitem{dong_study_2024}
W.-J. Dong, X.-Z. Yu, S.-Y. Ping, X.-T. Wu, G.~Wang, H.-Z. Huang, and Z.-W. Lin, ``Study of baryon number transport dynamics and strangeness conservation effects using \$\${\textbackslash}{Omega}\$\$-hadron correlations'', \href{http://dx.doi.org/10.1007/s41365-024-01464-8}{{\em Nuclear Science and Techniques} {\bfseries 35} (July, 2024) 120}. \url{https://doi.org/10.1007/s41365-024-01464-8}.

\bibitem{SciChinaJinS}
X.~H. Jin, J.~H. Chen, Z.~W. Lin, G.~L. Ma, Y.~G. Ma, and S.~Zhang, ``{Explore the QCD phase transition phenomena from a multiphase transport model}'', \href{http://dx.doi.org/10.1007/s11433-018-9272-4}{{\em Sci. China Phys. Mech. Astron.} {\bfseries 62} (2019) 11012}.

\bibitem{tang_investigating_2024}
S.-Y. Tang, L.~Zheng, X.-M. Zhang, and R.-Z. Wan, ``Investigating the elliptic anisotropy of identified particles in p–{Pb} collisions with a multi-phase transport model'', \href{http://dx.doi.org/10.1007/s41365-024-01387-4}{{\em Nuclear Science and Techniques} {\bfseries 35} (Mar., 2024) 32}. \url{https://doi.org/10.1007/s41365-024-01387-4}.

\bibitem{WangHai}
H.~Wang, J.~H. Chen, Y.~G. Ma, and S.~Zhang, ``{Charm hadron azimuthal angular correlations in Au+Au collisions at $\sqrt{{s}_{\mathrm{NN}}}=200$ GeV from parton scatterings}'', \href{http://dx.doi.org/10.1007/s41365-019-0706-z}{{\em Nucl. Sci. Tech.} {\bfseries 30} (2019) 185}.

\bibitem{wang_calculation_2024}
T.-T. Wang, Y.-G. Ma, and S.~Zhang, ``Calculation of momentum correlation functions between \${\textbackslash}ensuremath\{{\textbackslash}pi\}, {K}\$, and \$p\$ for several heavy-ion collision systems at \${\textbackslash}sqrt\{\{s\}\_\{{nn}\}\}=39\$ {GeV}'', \href{http://dx.doi.org/10.1103/PhysRevC.109.024912}{{\em Physical Review C} {\bfseries 109} (Feb., 2024) 024912}. \url{https://link.aps.org/doi/10.1103/PhysRevC.109.024912}.

\bibitem{ma_effects_2024}
Y.~Ma, ``Effects of $\alpha$-clustering structure on nuclear reaction and relativistic heavy-ion collisions'', \href{http://dx.doi.org/10.11889/j.0253-3219.2023.hjs.46.080001}{{\em Nuclear Techniques} {\bfseries 46} (Oct., 2024) 80001}. \url{https://www.hjs.sinap.ac.cn/zh/article/doi/10.11889/j.0253-3219.2023.hjs.46.080001/}.

\bibitem{chen_transport_2024}
Q.~Chen, G.~Ma, and J.~Chen, ``Transport model study of conserved charge fluctuations and {QCD} phase transition in heavy-ion collisions'', \href{http://dx.doi.org/10.11889/j.0253-3219.2023.hjs.46.040013}{{\em Nuclear Techniques} {\bfseries 46} (Oct., 2024) }. \url{https://www.sciengine.com/10.11889/j.0253-3219.2023.hjs.46.040013}. Publisher: Beijing Zhongke Journal Publising Co. Ltd.

\bibitem{HIJING-1}
X.-N. Wang and M.~Gyulassy, ``{HIJING: A Monte Carlo model for multiple jet production in pp, pA and AA collisions}'', \href{http://dx.doi.org/10.1103/PhysRevD.44.3501}{{\em Phys. Rev. D} {\bfseries 44} (1991) 3501--3516}.

\bibitem{HIJING-2}
M.~Gyulassy and X.~N. Wang, ``Hijing 1.0: A monte carlo program for parton and particle production in high energy hadronic and nuclear collisions'', \href{http://dx.doi.org/https://doi.org/10.1016/0010-4655(94)90057-4}{{\em Comp. Phys. Commun.} {\bfseries 83} (1994) 307 -- 331}.

\bibitem{ZPCModel}
B.~Zhang, ``Zpc 1.0.1: a parton cascade for ultrarelativistic heavy ion collisions'', \href{http://dx.doi.org/https://doi.org/10.1016/S0010-4655(98)00010-1}{{\em Comp. Phys. Commun.} {\bfseries 109} (1998) 193 -- 206}.

\bibitem{ARTModel}
B.~A. Li and C.~M. Ko, ``{Formation of superdense hadronic matter in high energy heavy-ion collisions}'', \href{http://dx.doi.org/10.1103/PhysRevC.52.2037}{{\em Phys. Rev. C} {\bfseries 52} (Oct, 1995) 2037--2063}. \url{https://doi.org/10.1103/PhysRevC.52.2037}.

\bibitem{HeQuarkCoalAMPT}
Y.~He and Z.-W. Lin, ``{Improved quark coalescence for a multi-phase transport model}'', \href{http://dx.doi.org/10.1103/physrevc.96.014910}{{\em Physical Review C} {\bfseries 96} (July, 2017) }. \url{http://dx.doi.org/10.1103/PhysRevC.96.014910}.

\bibitem{pOmegaCFinpp}
``{Unveiling the strong interaction among hadrons at the LHC}'', \href{http://dx.doi.org/10.1038/s41586-020-3001-6}{{\em Nature} {\bfseries 588} (Dec., 2020) 232–238}. \url{http://dx.doi.org/10.1038/s41586-020-3001-6}.

\bibitem{BEsourceHalo}
T.~Csörgő, B.~Lörstad, and J.~Zimányi, ``{Bose-Einstein correlations for systems with large halo}'', \href{http://dx.doi.org/10.1007/bf02907008}{{\em Zeitschrift für Physik C: Particles and Fields} {\bfseries 71} (July, 1996) 491–497}. \url{http://dx.doi.org/10.1007/BF02907008}.

\bibitem{CMS2020pipi13TeV}
A.~M. Sirunyan {\em et~al.}, ``{Bose-Einstein correlations of charged hadrons in proton-proton collisions at $\sqrt{s}$ = 13 TeV}'', \href{http://dx.doi.org/10.1007/jhep03(2020)014}{{\em Journal of High Energy Physics} {\bfseries 2020} (Mar., 2020) }. \url{http://dx.doi.org/10.1007/JHEP03(2020)014}.

\bibitem{ATLAS2015pipi}
G.~Aad {\em et~al.}, ``{Two-particle Bose–Einstein correlations in pp collisions at $\sqrt{s}$ = 0.9 and 7 TeV measured with the ATLAS detector}'', \href{http://dx.doi.org/10.1140/epjc/s10052-015-3644-x}{{\em The European Physical Journal C} {\bfseries 75} (Oct., 2015) }. \url{http://dx.doi.org/10.1140/epjc/s10052-015-3644-x}.

\bibitem{CMS2018pipi}
{\bfseries CMS} Collaboration, ``{Bose-Einstein correlations in pp, p--Pb, and Pb--Pb collisions at $\sqrt{s_{\rm{NN}}}$ = 0.9--7 TeV}'', \href{http://dx.doi.org/10.1103/PhysRevC.97.064912}{{\em Phys. Rev. C} {\bfseries 97} (Jun, 2018) 064912}. \url{https://link.aps.org/doi/10.1103/PhysRevC.97.064912}.

\bibitem{BEC_LevySource}
T.~Cs{\"o}rg{\H o}, S.~Hegyi, and W.~A. Zajc, ``{Bose-Einstein correlations for L{\'e}vy stable source distributions}'', \href{http://dx.doi.org/10.1140/epjc/s2004-01870-9}{{\em The European Physical Journal C - Particles and Fields} {\bfseries 36} (2004) 67--78}. \url{https://doi.org/10.1140/epjc/s2004-01870-9}.

\bibitem{pipiHBTinpp2011}
{\bfseries ALICE} Collaboration, K.~Aamodt {\em et~al.}, ``{Femtoscopy of pp collisions at $\sqrt{s}=0.9$ and 7 TeV at the LHC with two-pion Bose-Einstein correlations}'', \href{http://dx.doi.org/10.1103/PhysRevD.84.112004}{{\em Phys. Rev. D} {\bfseries 84} (Dec, 2011) 112004}. \url{https://link.aps.org/doi/10.1103/PhysRevD.84.112004}.

\bibitem{pipiHBTinPbPb2011}
K.~Aamodt {\em et~al.}, ``{Two-pion Bose–Einstein correlations in central Pb--Pb collisions at $\sqrt{s_{\rm{NN}}}$ = 2.76 TeV}'', \href{http://dx.doi.org/https://doi.org/10.1016/j.physletb.2010.12.053}{{\em Physics Letters B} {\bfseries 696} (2011) 328--337}. \url{https://www.sciencedirect.com/science/article/pii/S0370269310014565}.

\bibitem{Bhawanipd}
{\bfseries ALICE} Collaboration, S.~Acharya {\em et~al.}, ``{Exploring the Strong Interaction of Three-Body Systems at the LHC}'', \href{http://dx.doi.org/10.1103/PhysRevX.14.031051}{{\em Phys. Rev. X} {\bfseries 14} (Sep, 2024) 031051}. \url{https://link.aps.org/doi/10.1103/PhysRevX.14.031051}.

\bibitem{Dimi_CATS}
D.~L. Mihaylov, V.~Mantovani~Sarti, O.~W. Arnold, L.~Fabbietti, B.~Hohlweger, and A.~M. Mathis, ``{A femtoscopic correlation analysis tool using the Schrödinger equation (CATS)}'', \href{http://dx.doi.org/10.1140/epjc/s10052-018-5859-0}{{\em The European Physical Journal C} {\bfseries 78} (May, 2018) }.

\bibitem{BigBangToSmall}
K.~Yagi, T.~Hatsuda, and Y.~Miake, {\em {Quark-gluon plasma: From big bang to little bang}}, vol.~23.
\newblock 2005.

\bibitem{d_productionInAMPT}
Y.~Oh, Z.-W. Lin, and C.~M. Ko, ``{Deuteron production and elliptic flow in relativistic heavy ion collisions}'', \href{http://dx.doi.org/10.1103/PhysRevC.80.064902}{{\em Phys. Rev. C} {\bfseries 80} (Dec, 2009) 064902}. \url{https://link.aps.org/doi/10.1103/PhysRevC.80.064902}.

\end{thebibliography}\endgroup

\end{document}